\documentclass[sigconf]{aamas}

\usepackage{balance} 
\usepackage{tikz}
\usepackage{algorithm}
\usepackage[noend]{algpseudocode}
\usepackage{paralist}
\usepackage{subcaption}

\algdef{SE}[SUBALG]{Indent}{EndIndent}{}{\algorithmicend\ }%
\algtext*{Indent}
\algtext*{EndIndent}

\setcopyright{ifaamas}
\acmConference[AAMAS '23]{To appear at the 22nd International Conference
on Autonomous Agents and Multiagent Systems (AAMAS 2023)}{May 29 -- June 2, 2023}
{London, United Kingdom}{A.~Ricci, W.~Yeoh, N.~Agmon, B.~An (eds.)}
\copyrightyear{2023}
\acmYear{2023}
\acmDOI{}
\acmPrice{}
\acmISBN{}

\acmSubmissionID{1056}

\title[AAMAS-2023 Formatting Instructions]{A Comparison of New Swarm Task Allocation Algorithms in Unknown Environments with Varying Task Density}

\author{Grace Cai}
\affiliation{
  \institution{MIT}
  \city{Cambridge}
  \country{US}}
\email{gracecai@mit.edu}

\author{Noble Harasha}
\affiliation{
  \institution{MIT}
  \city{Cambridge}
  \country{US}}
\email{nharasha@mit.edu}

\author{Nancy Lynch}
\affiliation{
  \institution{MIT}
  \city{Cambridge}
  \country{US}}
\email{lynch@mit.edu}

\begin{abstract}
Task allocation is an important problem for robot swarms to solve, allowing agents to reduce task completion time by performing tasks in a distributed fashion. Existing task allocation algorithms often assume prior knowledge of task location and demand or fail to consider the effects of the geometric distribution of tasks on the completion time and communication cost of the algorithms. In this paper, we examine an environment where agents must explore and discover tasks with positive demand and successfully assign themselves to complete all such tasks. We first provide a new discrete general model for modeling swarms. Operating within this theoretical framework, we propose two new task allocation algorithms for initially unknown environments -- one based on N-site selection and the other on virtual pheromones. We analyze each algorithm separately and also evaluate the effectiveness of the two algorithms in dense vs. sparse task distributions. Compared to the Levy walk, which has been theorized to be optimal for foraging, our virtual pheromone inspired algorithm is much faster in sparse to medium task densities but is communication and agent intensive. Our site selection inspired algorithm also outperforms Levy walk in sparse task densities and is a less resource-intensive option than our virtual pheromone algorithm for this case. Because the performance of both algorithms relative to random walk is dependent on task density, our results shed light on how task density is important in choosing a task allocation algorithm in initially unknown environments.
\end{abstract}


\newcommand{\BibTeX}{\rm B\kern-.05em{\sc i\kern-.025em b}\kern-.08em\TeX}

\begin{document}


\pagestyle{fancy}
\fancyhead{}


\maketitle 


\section{Introduction} \label{sec:intro}

Robot swarms are simple, distributed units that are able to work together to achieve emergent collective behaviours \cite{fan2020review}. We contribute a general, theoretical framework to model these swarms, which can be leveraged, as will be done in this work, to implement various swarm algorithms. Swarm algorithms often draw inspiration from swarms in nature such as birds, ants, and bees \cite{pratt2005quorum, reynolds1987flocks, karaboga2009survey}. Swarm algorithms provide a scalable and fault-tolerant solution to problems such as search-and-rescue \cite{couceiro2017overview} and environmental monitoring \cite{duarte2016application}. One of the most well-studied swarm problems is task allocation \cite{gerkey2004formal}, which aims to assign agents to tasks in an optimal manner. Here, a \textit{task} refers to an abstract notion; a task is simply a location of interest in the environment which requires some action(s) by agents. This could be a food item in foraging, survivors in search-and-rescue, mines for robots to defuse, and more.

Many classes of task allocation algorithms assume that task locations and demand for agents are known, and try to optimize an assignment of agents to tasks \cite{bettinelli2020coalition, xu2015coalition, berman2009optimized}. However, in many applications, such as finding and defusing mines \cite{sahin2002swarm}, task information is not initially known. Algorithms which do consider task allocation in unknown environments \cite{harwell2018unknownta, pini2014ta} run limited testing on the effects of task density. However, the density of tasks in the environment affects relative algorithm performance. 

In this paper, we consider the problem of assigning agents to tasks with positive demand in an initially unknown environment. We assume each agent can only be assigned to one task. Within this setup, we contribute two new algorithms and compare them to the Levy Walk (RW), which is used in nature for foraging \cite{levy}. We also show how task density makes our different algorithms better suited towards different task environments.

The first algorithm, our house hunting task allocation algorithm (HHTA), is inspired by swarm house-hunting models \cite{reina2015design}. While the house hunting problem aims for agents to agree on one of many locations in the environment, the task allocation problem aims for agents to split themselves proportionally to task demand amongst all tasks in the environment. In our HHTA algorithm, agents use their starting location as a home base that they can return to after discovering tasks in the environment. The home base functions as a central point of communication and allows for agents to recruit each other to do tasks, serving the same function as the home nest in many swarm house hunting algorithms. 

The second algorithm is our propagation-based algorithm (PROP), which uses a regular grid of cheap, simple agents to propagate task demand information outwards to neighboring propagator agents. We assign a separate type of agent with more advanced computing powers to read the information and use it to probabilistically decide which task to head towards. The propagation of task demand information via cheap agents is inspired by virtual pheromones \cite{attygalle1985ant, barth2003dynamic}, a commonly used nature-inspired technique in swarm algorithms. 

By comparing both algorithms to the Levy flight, we show that it is harder for PROP to do well with very dense tasks, as a large influx of propagated information can confuse agents. Our other algorithm, HHTA, does worse when tasks are mid to high density because inter-agent communication about tasks is not worth it compared to a random walk, which is highly likely to encounter tasks quickly. However, it does better than RW when tasks are sparse as the cost of communication about task location is justified when tasks are harder to find. It is also less resource intensive compared to PROP. We also evaluate the effects of varying individual parameters within several task densities in order to better understand our new algorithms. 

Our results demonstrate how different task allocation algorithms do well in environments with different task density and invite further examination on the performance of other task allocation algorithms in different types of task environments. 

Section \ref{sec:bg} provides the inspiration for our two proposed algorithms, explaining house hunting and virtual pheromones in further depth. Section \ref{sec:model} describes our general formal model and then dives into the models for our two specific algorithms. Our simulation results and comparison between the two new algorithms in sparse and dense task environments can be found in Section \ref{sec:results}. Section \ref{sec:discussion} discusses our results, and Section \ref{sec:future} concludes the paper and provides ideas for future work. The full simulation code can be found at \cite{sourcecode}.

\section{Background} \label{sec:bg}
Task allocation is a well studied problem and has been classified into many subproblems. Per the taxonomy defined in \cite{gerkey2004formal} our task allocation problem is of the single-task agents, multi-robot tasks variety, which means that agents can only do one task at a time, but tasks may require multiple agents. 

When task demands and locations are known, this problem becomes a coalition formation problem, where we wish to form agents into groups that are best suited to do each task. This problem can be thought of as a set partitioning problem \cite{gerkey2004formal}, and adaptations to distributed swarms have been proposed \cite{bettinelli2020coalition, xu2015coalition}.

Other strategies for when tasks are fixed at known locations model tasks as a graph where agents can travel between edges \cite{berman2009optimized, hsieh2008biologically, halasz2007dynamic}. These algorithms optimize for a flow rate between edges in the graph so that agents can satisfy all task demands quickly. Another strategy in this case, based on Optimal Mass Transport \cite{solomon2018optimal}, is to treat the tasks with demands as sinks and the tasks with agents as sources in a min cost flow problem. However, both strategies require prior knowledge of task locations. 

Our problem differs from coalition formation and the graph-based task allocation problems because we are assuming that agents have no initial knowledge of task location or demands. In this case, we want to discover tasks and communicate information about them as quick as possible so that agents can satisfy all task demands.

One solution to task allocation in an environment with unknown tasks is to have agents form local clusters and run Optimal Mass Transport locally \cite{zhang2007adaptive}. Other task allocation algorithms, such as auction-based algorithms, perform a similar type of agent clustering to assign tasks \cite{hoeing2007auction}. Our two algorithms by contrast are fully distributed and computationally simple, without the need for grouping to locally run a complex centralized algorithm. This allows us to save the time needed to form agent clusters and allows agents to be cheaper to implement due to low computation cost.

\subsection{Levy Flight}
The Levy flight is a random walk that has been observed in foraging animals and adapted to swarm algorithms \cite{levy, levy2} as well. The Levy Flight has shown to be an optimal forgaing algorithm, which is very relevant to the situation in which task locations and demands are unknown. As such, we will be using this random walk as a baseline to compare against for our two new algorithms. 

\subsection{House Hunting}
Several ant species engage in a house-hunting behaviour when their home nest is destroyed \cite{pratt2005quorum, pratt2005behavioral}. First, ants explore nearby for nest sites. If a site is found, the ant waits a period of time inversely proportional to the site quality before returning to the home nest to lead others to the new nest. This process of recruitment is known as forward tandem running (FTR). Once the encounter rate of other ants in the candidate nest reaches a critical threshold known as the quorum threshold, ants switch to carrying members of the colony to the new nest. This carrying behaviour is 3 times faster than FTR and accelerates the move to the new site \cite{pratt2005quorum}. 

Ant house hunting has inspired the corresponding swarm problem of N-site selection \cite{valentini2017best}, where agents must choose the highest quality site from N initially unknown candidates. One common N-site selection model has agents transition between four main states: Uncommitted Interactive, Uncommitted Latent, Favoring Interactive, and Favoring Latent \cite{reina2015design}. Some works also include a fifth Committed state \cite{khurana2020quorum, cody2017evaluation, cai2019urgency}. In this type of model, Uncommitted Interactive agents explore the arena for new sites, while Uncommitted Latent agents stay in the home nest. Once an Uncommitted Interactive agent discovers a site, it can decide to favor the site. Favoring agents can be interactive, meaning they return to the home nest to recruit other favoring agents, or latent, meaning they stay in their favored site to build up quorum. Lastly, if agents detect a sufficient number of others in a new candidate site, they can transition into the committed state to finalize their decision.  

Task allocation can be thought of as an extension to the house hunting problem, where instead of trying to send all agents to one location, we want to send agents to multiple locations according to the demand at each one. This idea has been used in Berman \cite{berman2009optimized} and Halasz's \cite{halasz2007dynamic} work to develop task allocation algorithms for a known graph of tasks where agents can traverse along the edges. We extend this idea further by using inspiration from site selection algorithms to develop our novel HHTA algorithm, in which, unlike \cite{berman2009optimized, halasz2007dynamic}, task locations are initially unknown. In HHTA, agents use a home nest which functions as a location for recruiting other agents to tasks and communicating with other agents. The four main states of the HHTA algorithm share parallels to the Uncommitted Interactive, Uncommitted Latent, Favoring Active, and Committed states described above which are further explained in Section \ref{subsec:hhta}.

\subsection{Virtual Pheromones and Potential Fields}
Ants leave pheromones in their environment when foraging to guide other ants to any
discovered food sources \cite{attygalle1985ant}. This strategy of leaving information in the environment has inspired swarms to implement virtual pheromones (pheromones represented by computational data instead of chemical signals). For example, \cite{barth2003dynamic} used physically deployable beacons that robots could leave in the environment to store information in, \cite{lumoses2018foraging} simulated pheromone trails by leveraging depots to store target-rich locations (pheromone waypoints) found by other robots, and \cite{o2005gnats} set up a virtual pheromone approach with a pre-deployed network of beacons that acted as a grid of locations to leave information in. One cheap way to implement virtual pheromones is using wireless sensor motes to store and propagate information \cite{russell2015swarm}.

Pheromones are frequently used in conjunction with potential fields or particle swarm optimization techniques. Potential field algorithms model objects in the environment as either positive charges or negative charges, with agents experiencing attraction or repulsion from the objects based on the electric force between them. Particle swarm optimization \cite{poli2007particle} follows a similar physics approach, except the attractive and repulsive forces were based on springs as opposed to charges. These techniques are employed in navigation tasks, where potential fields and pheromones can work together to guide robots around obstacles and towards a target in space \cite{parunak2002digital}. Pheromones are also employed in foraging tasks to help robots efficiently find what they are foraging for \cite{hoff2010two}.

We apply the ideas of virtual pheromones in our novel PROP algorithm, which uses simple mote-like agents to leave task information in the environment. Task-performing robots use this information when searching for tasks in the task allocation process. The use of virtual pheromones allows us to easily notify task-performing robots of nearby tasks. We also use potential fields as inspiration for how a robot’s motion should be influenced when it learns of multiple potential tasks through pheromones in the environment. Robots are more attracted to tasks with higher demand and tasks that are closer to their current location, so tasks can be thought of like charges which robots can feel the force of. 

\section{Model} \label{sec:model}
We first describe our new discrete general model for modeling swarms. Then we discuss the individual restrictions, parameters, and agent algorithms needed for task allocation. Figure \ref{fig:paramtable} is provided as a lookup table for the parameter notation used in defining this theoretical framework (and our two task allocation algorithms).

\begin{figure}
    \centering
    \begin{tabular}{ |p{1.5cm}||p{5cm}|  }
     \hline
     Notation & Parameter Description\\
     \hline
     \hline
     $M$, $N$ & Grid dimensions, $M \times N$ \\
     \hline
     $I$ & Agent influence radius \\
     \hline
     $\alpha$ & Agent transition function \\
     \hline
     $T$ & Number of tasks \\
     \hline
     $t_i$ & Task $i$ \\
     \hline
     ${t_i}^{rd}$ & Residual demand of $t_i$ \\
     \hline
     HHTA & House hunting task allocation algorithm \\
     \hline
     PROP & Task propagation algorithm \\
     \hline
     RW & Levy random walk \\
     \hline
     $P_c$ & Base probability of committing to a task upon arrival (for HHTA) \\
     \hline
     $P_e$ & Expected fraction of exploring agents (for HHTA) \\
     \hline
     $r_m$ & Recruiting agents' message rate (for HHTA) \\
     \hline
     $\mathcal{M_T}$ & Propagator agent's known task demand information (for PROP) \\
     \hline
     $d_p$ & Maximum propagation radius (for PROP) \\
     \hline
     $r_p$ & Propagation timeout in integer rounds (for PROP) \\
     \hline
     $L$ & $1/(M+N)$ \\
     \hline
    \end{tabular}
    \caption{Summary table of notation for relevant parameters.}
    \label{fig:paramtable}
\end{figure}

\subsection{General Model}
We assume a finite set $R$ of agents, with a state set $SR$ of potential states. Agents move on a discrete rectangular grid of size $M \times N$, formally modelled as directed graph $G = (V,E)$ with $|V| = MN$. Edges are  bidirectional, and we also include a self-loop at each vertex. Vertices are indexed as $(x, y)$, where $0 \leq x \leq M-1$, $0 \leq y \leq N-1$. Each vertex also has a state set $SV$ of potential states.

\textit{Local Configurations:} A \textit{local configuration} $C'(v)$ captures the contents of vertex $v$. It is a triple $(sv, myagents, srmap)$, where $sv \in SV$ is the vertex state of $v$, $myagents \subseteq R$ is the set of agents at $v$, and $srmap: myagents \rightarrow SR$ assigns an agent state to each agent at $v$.


\textit{Local Transitions:} The transition of a vertex $v$ may be influenced by the local configurations of nearby vertices. We define an \textbf{influence radius} $I$, which is the same for all vertices, to mean that vertex indexed at $(x,y)$ is influenced by all valid vertices $\{(a,b) | a \in [x-I, x+I], b\in [y-I, y+I]\}$, where $a$ and $b$ are integers. 
We can use this influence radius to create a local mapping $M_v$ from local coordinates to the neighboring local configurations. For a vertex $v$ at location $(x,y)$, we produce $M_v$ such that $M_v(a,b) \rightarrow C'(w)$ where $w$ is the vertex located at $(x+a, y+b)$ and $ -I < a,b < I$. This influence radius is representative of a sensing and communication radius. Agents can use all information from vertices within the influence radius to make decisions.

We have a local transition function $\delta$, which maps all the information associated with a vertex and its influence radius at one time to new information that can be associated with the vertex and the agents at that vertex for the following time.

Formally, for a vertex $v$, $\delta$ probabilistically maps $M_v$ to a quadruple of the form $(sv_1, myagents, srmap_1, dirmap_1)$, where $sv_1 \in SV$ is the new state of the vertex, $srmap_1: myagents \rightarrow SR$ is the new agent state mapping for agents at the vertex, and $dirmap_1: myagents \rightarrow \{R,L,U,D,S\}$ gives directions of motion for agents currently at the vertex. Note that $R$, $L$, $U$, and $D$ mean right, left, up, and down respectively, and $S$ means to stay at the vertex. The local transition function $\delta$ is further broken down into two phases as follows.

\textit{Phase One:} Each agent in vertex $v$ uses the same transition function $\alpha$, which probabilistically maps the agent's state $sr \in SR$, location $(x,y)$, and the mapping $M_v$ to a new suggested vertex state $sv'$, agent state $sr'$, and direction of motion $d \in \{R, L, U, D, S\}$. We can think of $\alpha$ as an agent state machine model. 

\textit{Phase Two:} Since agents may suggest conflicting new vertex states, a rule $Q$ is used to select one final vertex state. The rule also determines for each agent whether they may transition to state $sr'$ and direction of motion $d$ or whether they must stay at the same location with original state $sr$.

\textit{Probabilistic Execution: }
\sloppy
The system operates by probablistically transitioning all vertices $v$ for an infinite number of rounds. During each round, for each vertex $v$, we obtain the mapping $M_v$ which contains the local configurations of all vertices in its influence radius. We then apply $\delta$ to $M_v$ to transition vertex $v$ and all agents at vertex $v$. For each vertex $v$ we now have $(sv_v, myagents_v, srmap_v, dirmap_v)$ returned from $\delta$.

For each $v$, we take $dirmap_v$, which specifies the direction of motion for each agent and use it to map all agents to their new vertices. For each vertex $v$, its new local configuration is just the new vertex state $sv_v$, the new set of agents at the vertex, and the $srmap$ mapping from agents to their new agent states. 

\vspace{-0.3em}

\subsection{Task Allocation Problem Definition}
Consider $T$ tasks $[t_0 \dots t_{T-1}]$ arranged at a subset of vertices in our general model, with at most one task at each vertex. Specifically, the task locations can be described as $l = [(x_0, y_0), \dots ,(x_{T-1}, y_{T-1})]$, where $l_i = (x_i, y_i)$ is the vertex location of task $t_i$ and $i \neq j \rightarrow l_i  \neq l_j$ (each task has a distinct location). We wish to distribute agents among the tasks to achieve a certain distribution $a = [a_0, . . . , a_{T-1}]$ where $a_i$ represents the number of agents doing task $i$ and $\sum a_i = kR \leq R$ (meaning $k$\% of all agents is enough to complete all the tasks). 

We assume that when an agent senses a task within its influence radius, it is able to detect the demand of that task. Since agents can also detect how many agents are at the task, they can use this information to compute the \textit{residual demand}, defined as the difference between the task demand and the number of agents already at the task. We denote the residual demand at task $i$ by $t_i^{rd}$. We assume that the desired task distribution does not change over time, and that the task is complex enough that each agent can only do one task over the course of the algorithm.

In order to properly represent tasks in both of our algorithms, the vertex state set $SV$ contains the following variables: \texttt{is\_task}, whether the vertex is a task; \texttt{demand}, the task demand if the vertex is a task; \texttt{residual\_demand}, the residual demand if the vertex is a task; \texttt{task\_location}, the $x,y$ coordinates of the vertex if it is a task.

We go into more detail on the agent states and transitions for our two algorithms in Sections \ref{subsec:hhta} and \ref{subsec:prop}. One other detail to note about task allocation is that in phase two of $\delta$, we reconcile conflicting proposed vertex states. This shows up in task allocation when multiple agents attempt to claim the same task. When this happens, if there are $s$ agents trying to claim the task but only $rd < s$ residual demand, then only $rd$ agents are allowed to transition their state to having claimed the task (these $rd$ agents are chosen arbitrarily). Otherwise, if $rd > s$, all agents will be allowed to claim the task. 

\subsection{House Hunting Task Allocation Algorithm} \label{subsec:hhta}
In our house-hunting inspired algorithm (HHTA), agents start out at a square home location with lower left corner $(x^1_h, y^1_h)$ and upper right corner $(x^2_h, y^2_h)$. Call the set of home vertices $\mathcal{H}$. We assume that $\forall i, l_i \notin \mathcal{H}$, meaning no tasks are located at the home location. In HHTA, the vertex state set $SV$ needs the additional variable \texttt{is\_home}, indicating whether the vertex is a home vertex or not. 

Agents can be in one of four core states: Home (H), Exploring (E), Recruiting (R), or Committed (C). Home agents wait in home nest for news of tasks. Exploring agents explore the arena for tasks. Home agents have a $P_E = \frac{L*P_e}{1-P_e}$ chance of converting to exploring agents, and exploring agents have a $P_H = L$ chance of converting to home agents, where $L$ is defined as $1/(M+N)$ and $P_e$ is the expected fraction of exploring agents. The transitions between H and E agents indicate that agents are expected to explore for $M+N$ time steps (enough to reach the corners of the grid) before returning home. It also ensures that the expected fraction of E agents out of the total number of E and H agents is $P_e$. The factor of $L$ is inspired by house hunting algorithms, where $L$ is defined as the inverse of the average site round trip so that exploring agents will have enough time to reach candidate sites before returning home. 

An exploring agent has a $P_{t_i}$ chance of finding task $i$. Once it finds task $i$, it has a $c = \max(P_c, 1/t_i^{rd})$ chance of becoming a Committed agent, and a $1-c$ chance of becoming a Recruiting agent. Here $P_c$ is the base probability of committing, and $1/t_i^{rd}$ makes it so that at low residual demands, agents have a higher chance of committing to the task right away. If a task has residual demand $1$, for instance, any agent which discovers it will commit to the task right away, completing the task instead of trying to recruit others for it. 

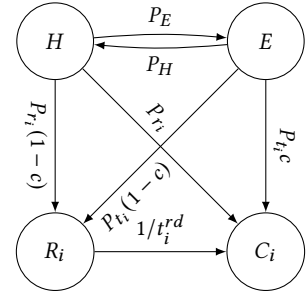
\begin{figure}
    \centering
    \begin{tikzpicture}[>=latex, scale=0.7]
    \fill[fill=white]
    (0,4) node[circle,inner sep=0pt,draw, text width = 1cm, text centered](H) {$H$}
 -- (4, 4) node[circle, inner sep = 0pt, draw, text width = 1cm, text centered](E){$E$}
 -- (4, 0) node[circle, inner sep = 0pt, draw, text width = 1cm, text centered](CI){$C_i$}
 -- (0, 0)node[circle, inner sep = 0pt, draw, text width = 1cm, text centered](RI){$R_i$};
\begin{scope}[every node/.style={scale=.95}]
 \path[->, sloped]
(H) edge[bend left=5] node[above] {$P_E$} (E)
(E) edge[bend left=5] node[below]{$P_H$} (H)
(E) edge[] node[above] {$P_{t_i}c$} (CI)
(E) edge[] node[below, anchor = north east] {$P_{t_i}(1-c)$} (RI)
(H) edge[] node[above, anchor = south east] {$P_{r_i}$} (CI)
(H) edge[] node[below] {$P_{r_i}(1-c)$} (RI)
(RI) edge[] node[above] {$1/t_i^{rd}$} (CI)
;
\end{scope}
\end{tikzpicture}
    \caption{State model of the four core states. The subscript $i$ denotes that an agent is recruiting for or committed to task $i$.}
    \label{fig:hh-transitions}
\end{figure}

Committed agents have fully committed to a task and stay at that task. The Committed state is similar to the Committed state in house hunting, where agents have decided on a new nest site and have moved to it. Recruiting agents head back to the home nest to tell Home agents about the task they have found. Agents recruiting for site $i$ have a $1/t_i^{rd}$ chance to stop recruiting and become committed to task $i$. Recruiting agents have a $r_m$ chance of sending a message to each agent within their influence radius at each time step, where $r_m$ is the message rate. Therefore, a Home agent has an $P_{r_i} = I_{1-r_m}(R_{t_i}-1, 2)$ chance of receiving at least one recruiting message for task $i$. Here, $R_{t_i}$ is the number of agents recruiting for task $i$ that are within sensing radius, and $I$ is the regularized incomplete beta function. If a Home agent receives a message from a recruiting agent, it has a $P_c$ chance of committing to the task and heading towards it, and a $1-P_c$ chance of recruiting for the task. Note that the residual demand information for C and R agents may become stale as more agents commit to tasks. A diagram of the transitions between these core states can be found in Figure \ref{fig:hh-transitions}.

In order to execute the core state transitions, the agent state set $SR$ comprises of the following variables: \texttt{core\_state}, which can be H, E, R or C; \texttt{id}, the agent id, taking on values from $0 \dots |R|-1$; \texttt{L}, defined as $L = 1/(M+N)$; \texttt{P\_commit}, the probability $P_c$; \texttt{P\_explore}, the probability $P_e$; \texttt{message\_rate}, the message rate $r_m$; \texttt{angle}, the agent's current angle of travel; \texttt{starting\_point}, a random walk parameter tracking where the agent started from; \texttt{travel\_distance}, the length of the current leg of the random walk; \texttt{destination\_task}, the agent's destination if they have just found a task or are headed towards their committed task; \texttt{home\_destination}, the agent's destination if they are headed to a home vertex; \texttt{recruitment\_task}, the task an agent is recruiting for; and \texttt{committed\_task}, the task an agent has committed to.

The agent transition function $\alpha$ uses these state variables to implement the transitions between the four core states. 

A pseudocode example of how $\alpha$ looks like for Recruiting agents can be seen in Algorithm \ref{alg:hh-code}. In this example, we first check if the recruiting agent has reached the home nest to recruit yet. If they have not, they keep heading towards the home nest by stepping one step in that direction. If they have reached the home nest, they have a $1/t_i^{rd}$ chance of transitioning to the Committed core state. Otherwise, they remain in the Recruiting state. 

\begin{algorithm}
\caption{Agent transition function $\alpha$ for a Recruiting agent with state $s$ at vertex $v$ with coordinates $(x,y)$}\label{euclid}
\begin{algorithmic}[1]
\Procedure{generate\_transition}{local\_vertex\_mapping}
\State $new\_agent\_state \gets s$
\If {$s.\texttt{home\_destination is not None}$}
\State $new\_direction \gets$ 
\Indent \State $\texttt{dir\_from\_dest(}s.\texttt{home\_destination, x, y)}$ \EndIndent
\State $new\_location \gets$
\Indent \State $\texttt{coords\_from\_dir(x,y,}new\_direction\texttt{)}$ \EndIndent
\If {$\texttt{within\_home(}new\_location\texttt{)}$}
\State $new\_agent\_state.\texttt{home\_destination} \gets \texttt{None}$
\State \Return $v.\texttt{state}$, $new\_agent\_state, new\_direction$
\EndIf
\EndIf
\State $committed\_chance \gets$ 
\Indent \State $1/s.\texttt{recruitment\_task.residual\_demand}$ \EndIndent
\If {$\texttt{random\_float\_from(0,1) < }committed\_chance$}
\State $new\_agent\_state.\texttt{core\_state} \gets \texttt{Committed}$
\State $new\_agent\_state.\texttt{destination\_task} \gets$
\Indent \State $new\_agent\_state.\texttt{recruitment\_task}$ \EndIndent
\State $new\_agent\_state.\texttt{recruitment\_task} \gets \texttt{None}$
\State \Return $v.\texttt{state}, new\_agent\_state, \texttt{S}$
\EndIf
\State \Return $self.\texttt{location.state}, new\_agent\_state, \texttt{S}$
\EndProcedure
\end{algorithmic}
\label{alg:hh-code}
\end{algorithm}

\subsection{Task Propagation Algorithm} \label{subsec:prop}
In our task propagation algorithm (PROP), we distinguish between two types of agents -- $MN$ propagators and $F$ followers. Propagators are simple, mote-like agents. One of them is assigned to each vertex to allow vertices to propagate task information to each other. Followers are more advanced agents which are able to perform the tasks in the task allocation problem. Followers follow the signals left by propogators in order to find tasks.

Similarly to HHTA, all agents are initially deployed at a rectangular home location with lower left corner $(x^1_h, y^1_h)$ and upper right corner $(x^2_h, y^2_h)$. However, agents in PROP do not utilize this home location after starting the algorithm. First, all $MN$ propagators travel to the vertex which they are assigned to, taking $\frac{M+N}{2}$ time for all agents to reach their assigned vertex. 

\begin{figure}
\centering
\begin{subfigure}[t]{.47\linewidth}
\includegraphics[width=\linewidth]{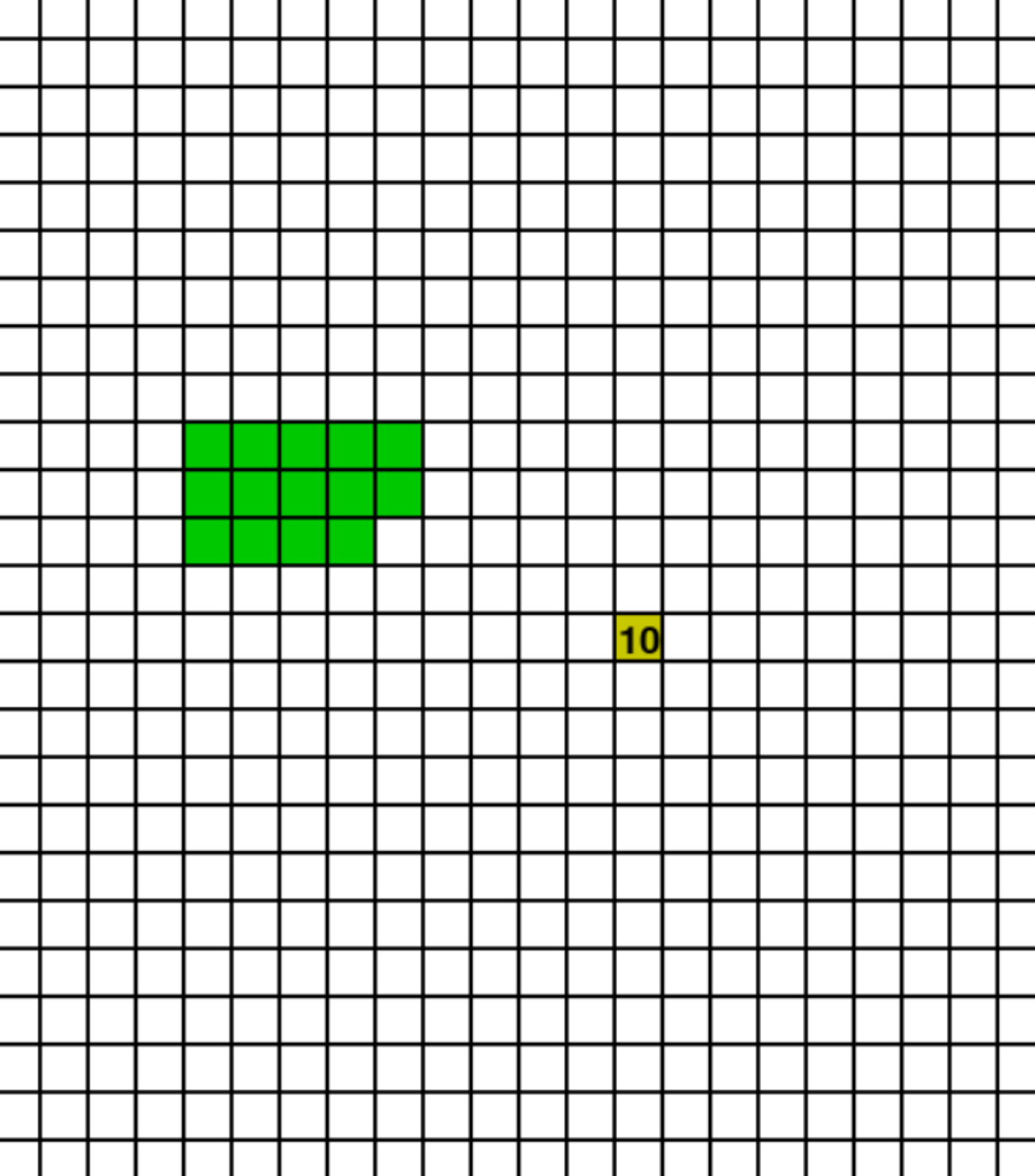}
\caption{Initial state}\label{fig:sub1}
\end{subfigure}
\begin{subfigure}[t]{.47\linewidth}
\includegraphics[width=\linewidth]{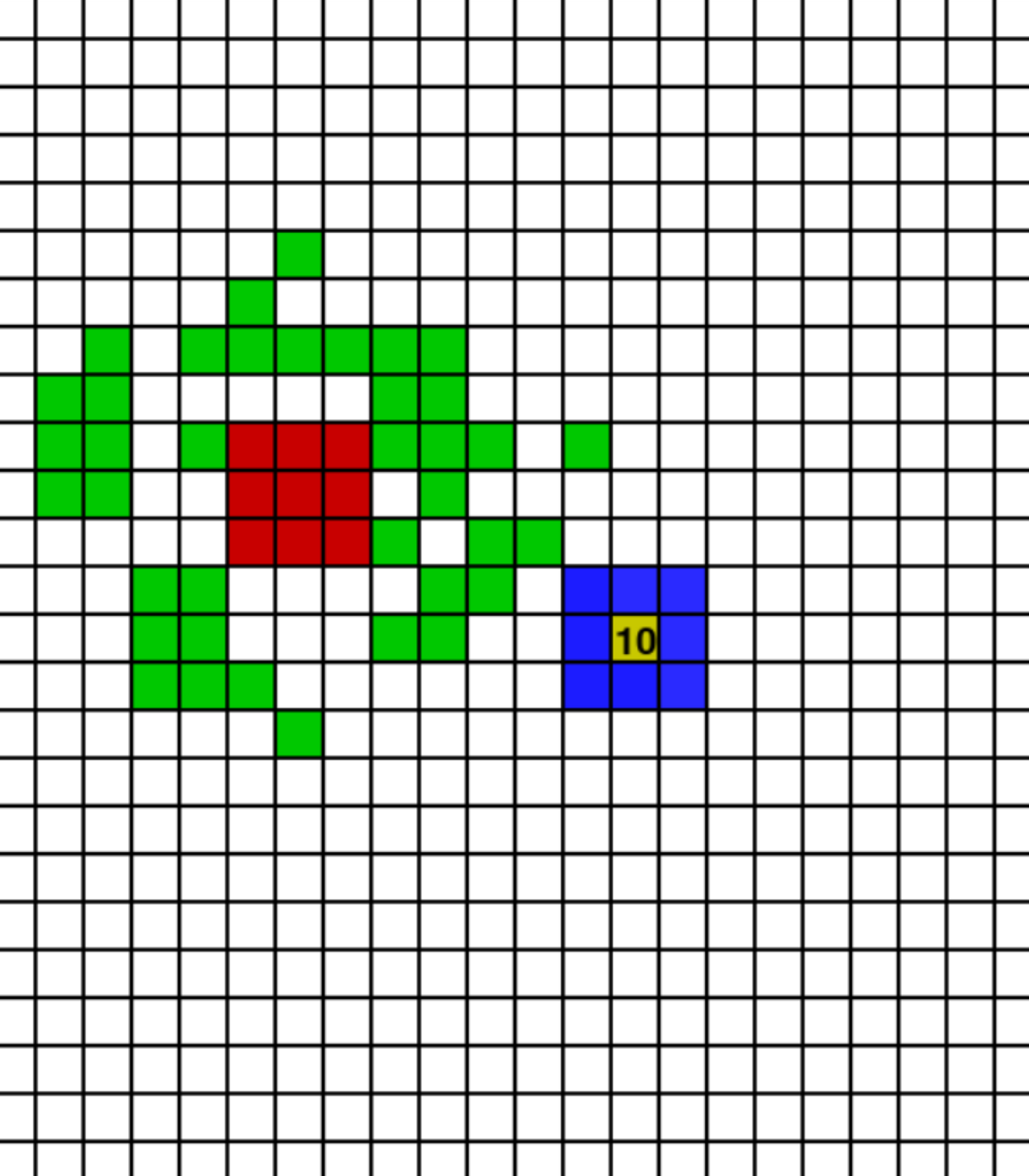}
\caption{Propagator agent at task propagates demand info to its influence radius of 1}\label{fig:sub2}
\end{subfigure}

\begin{subfigure}[t]{.47\linewidth}
\includegraphics[width=\linewidth]{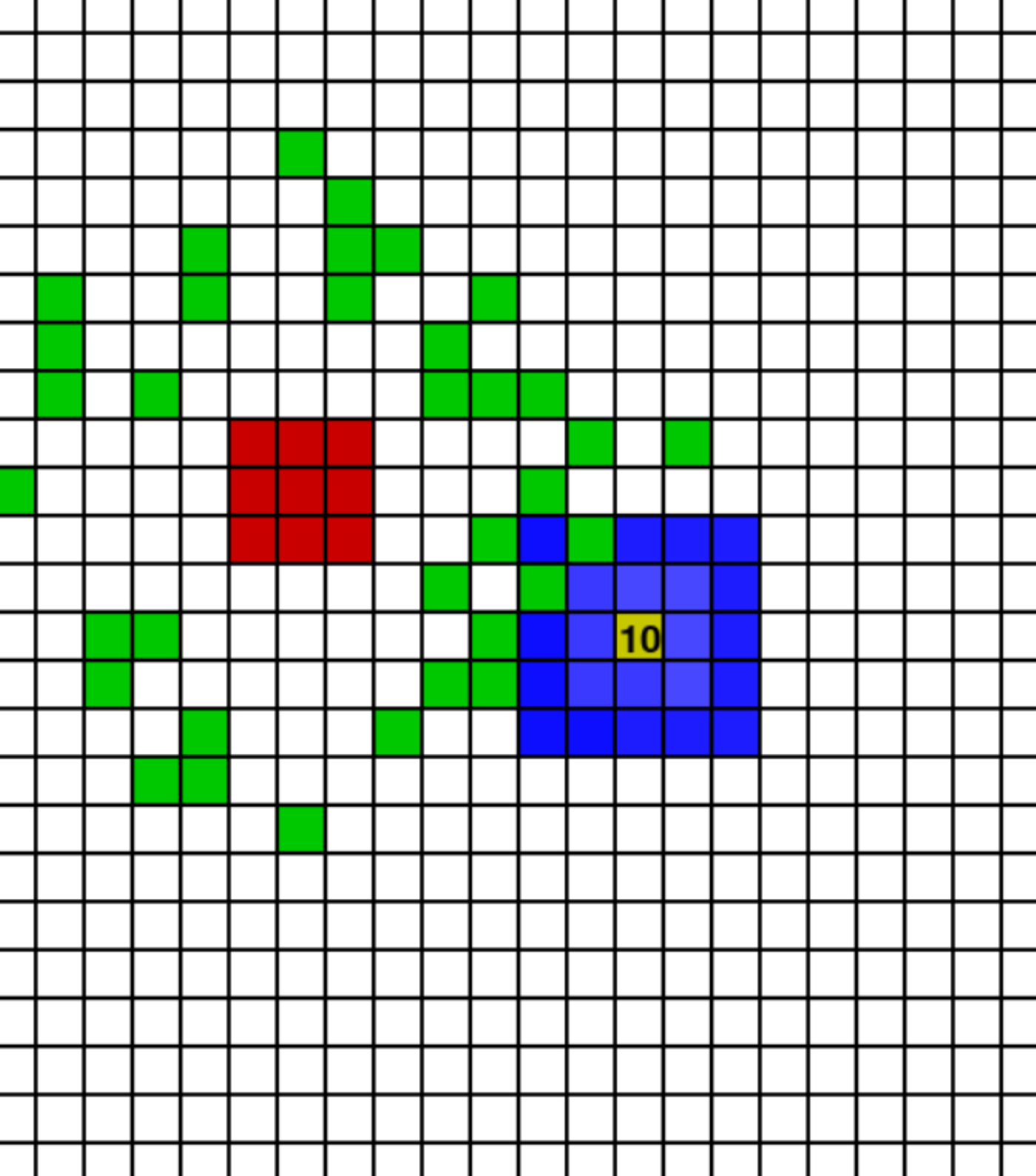}
\caption{$r_p$ rounds later}\label{fig:sub3}
\end{subfigure}
\begin{subfigure}[t]{.47\linewidth}
\includegraphics[width=\linewidth]{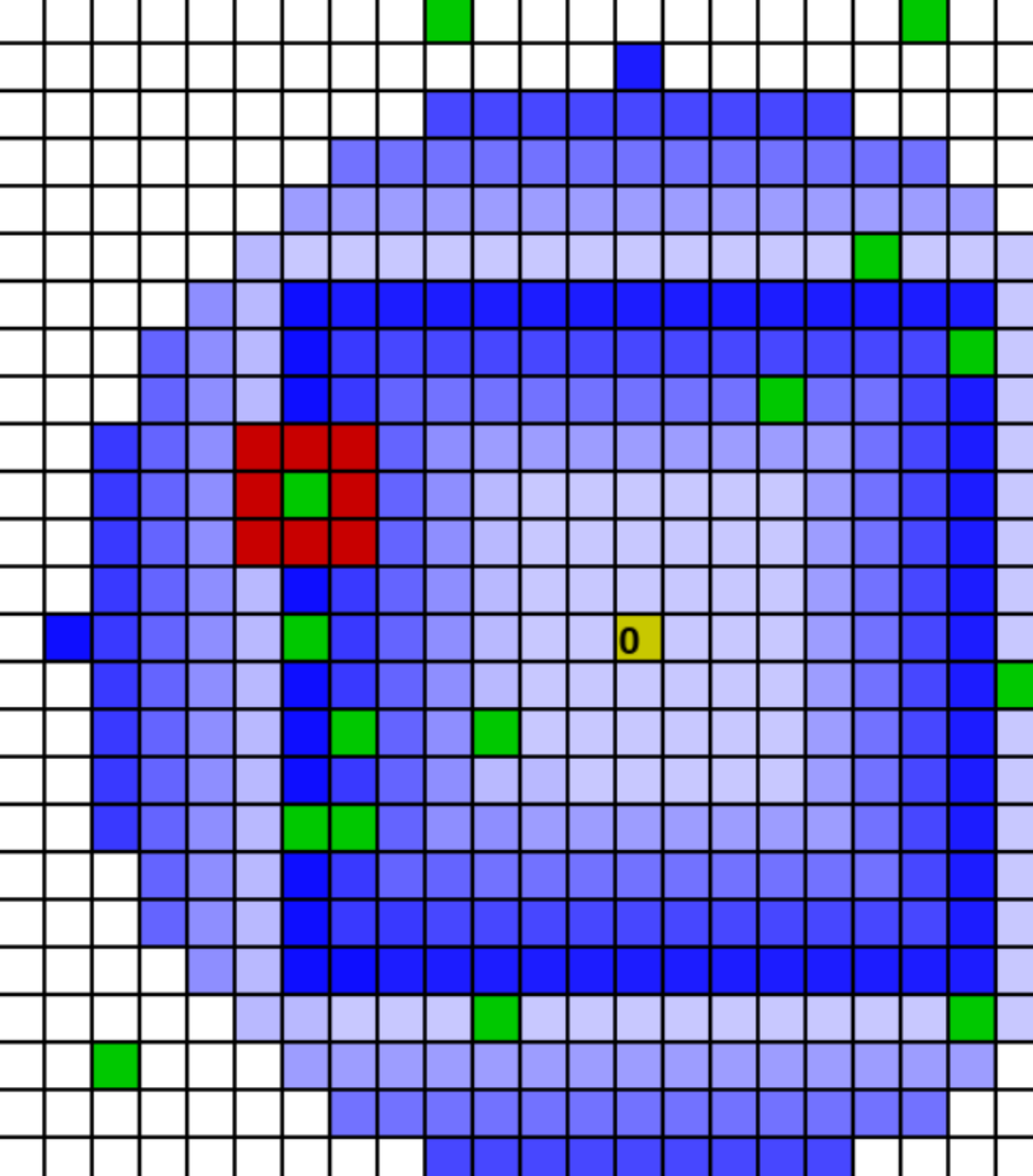}
\caption{Propagation bounded by $d_p$, radius of circle here}\label{fig:sub4}
\end{subfigure}

\begin{subfigure}[t]{.47\linewidth}
\includegraphics[width=\linewidth]{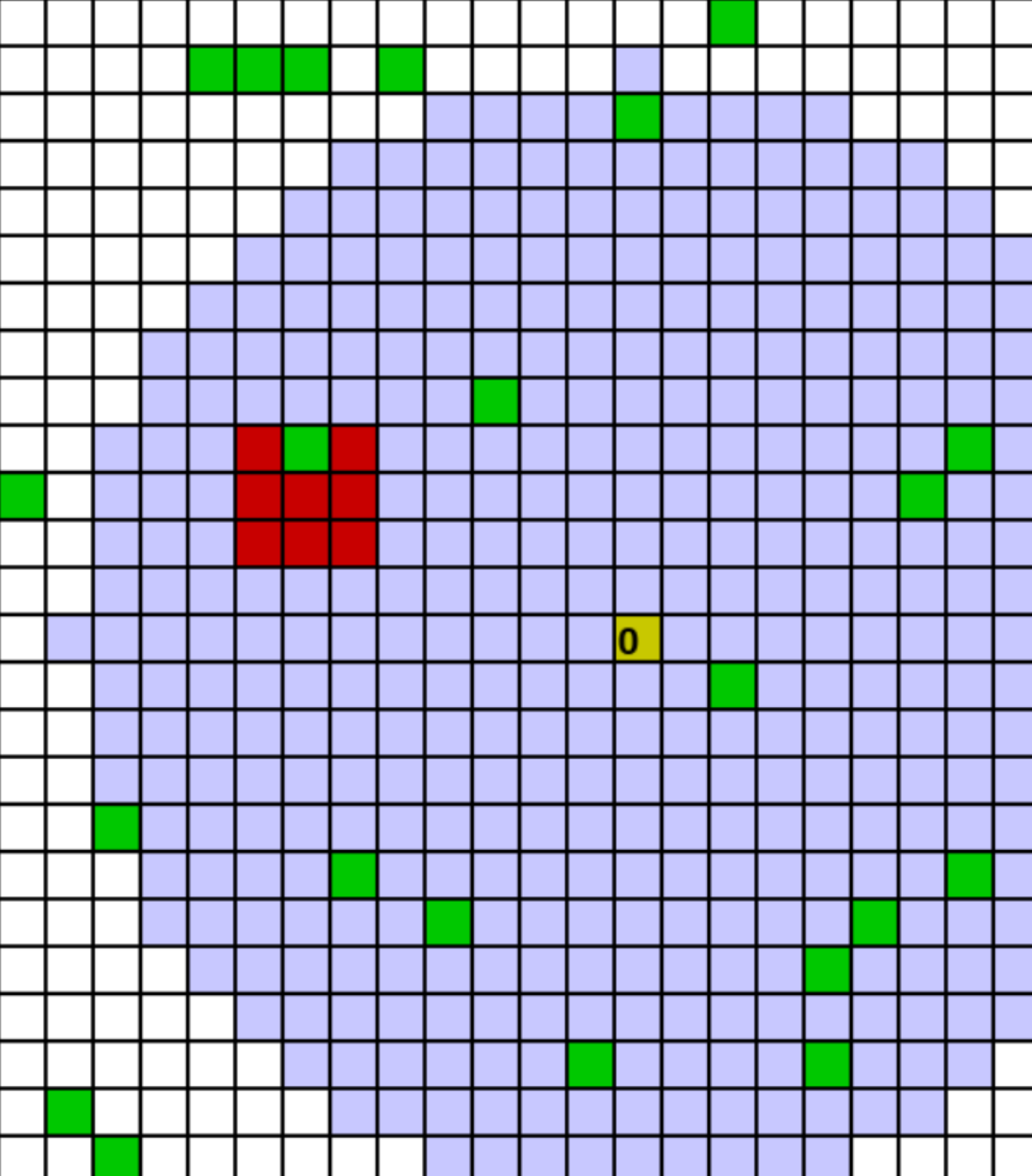}
\caption{No new task demand info to propagate}\label{fig:sub5}
\end{subfigure}
\caption{A simulated example---(a) to (e) chronologically---of how task demand information is propagated by propagator agents. Each square in the grid is a vertex, with edges between adjacent squares. Red denotes home vertices, green vertices have follower agents, the numbered vertex is a task with that value as its current residual demand, and blue vertices have propagator agents containing task demand information, with darker blues denoting newer information.}
\label{fig:propsim}
\end{figure}

Each propagator has an influence radius of 1 and also stores in their state a mapping $\mathcal{M}_T$ from task locations $l_i$ to residual demands $t_i^{rd'}$, representing that they have heard that task $i$ at location $l_i$ has residual demand $t_i^{rd'}$. After all propagators are in place, propagators that are at a task $t_i$ spread the tuple $(t_i^{rd'}, l_i)$ to all other propagators in their influence radius. Every $r_p$ time steps, a propagator takes all new task information (if it has new information it did not already propagate) it has received and spreads that information to all other propagators in its influence radius with the following conditions: information about task $i$ can only be spread to agents whose assigned vertex $v$ is located within the bounds $[x_i-I,x_i+I]$ for the x coordinate and $[y_i-I, y_i+I]$ for the y coordinate, and the Euclidian distance between $t_i$ and $v$ must be less than or equal to $d_p$. Here, $r_p$ is the integer propagation timeout and $d_p$ is the maximum propagation radius. Figure \ref{fig:propsim} shows an example of this propagation process, with propagator agents spreading the demand information of a single task throughout the graph.

Because the residual demand of a task changes over time, the propagator at task $i$ will have to send new information whenever the residual demand decreases. When a propagator which already has task information $\mathcal{M_T}(l_i) \rightarrow t_i^{rd'}$ receives new information about a task $(t_i^{rd''}, l_i)$, it updates the task information for task $i$ to be $\mathcal{M_T}(l_i) \rightarrow min(t_i^{rd'}, t_i^{rd''})$ in order to have the most up-to-date information. Since the residual demand of a task is always decreasing as more and more agents join the task, we know the smaller residual demand is the more accurate one. 

After all propagators have reached their assigned vertex, followers try to use the information of propagators in order to find tasks to head towards. At every time step, a follower first checks the vertices within its influence radius for a task with non-zero residual demand, and starts moving towards that task if it exists. If no task is found in its influence radius, a follower located at $(x,y)$ looks at the propagator assigned to location $(x,y)$ in order to get information about potential task locations it could head towards. It compiles all non-zero residual demands into the resulting mapping $M_F$, which maps from task locations $l_i$ to residual demands $t_i^{rd'}$. If $M_F$ is non-empty (there is at least one task location with non-zero residual demand) then the probability that a follower located at $(x,y)$ heads towards task location $l_i \in M_F$ is:
\begin{equation}
    \frac{\frac{M_F(l_i)}{L_2(l_i, (x,y))^2}}{\sum\limits_{l_j \in D(M_F)}\frac{M_F(l_j)}{L_2(l_j, (x,y))^2}}
\end{equation}

This means that the probability of a follower heading towards a task has an inverse square relationship with $L_2$ distance between the task location and the agent's location, and is also weighted by the residual demand of the task itself. This equation is determined so that agents are less likely to travel to tasks that are further away from them, but more likely to travel to a task if it has higher residual demand. If the mapping $M_F$ is empty (the agent has no task information), it takes a random step in one direction $\{L, D, R, U\}$ (following a Levy flight random walk) in order to explore. 

Once a follower agent reaches a task with non-zero residual demand, it stays there indefinitely, "completing the task" and decrementing the task's residual demand by one. 

In order to execute the algorithm, the agent state set contains the following variables: \texttt{type}, the type of agent, which can be `propagator' or `follower' and \texttt{id}, the agent id, which takes on values from $0 \dots |MN + F|-1$. The following additional variables are in SR and are only used by propagator agents: \texttt{task\_info}, the mapping $\mathcal{M_T}$; \texttt{propagation\_rate}, the propagation timeout $r_p$; and \texttt{propagation\_ctr}, the number of rounds since an agent last propagated task information. Lastly, the variables in SR used only by follower agents are: \texttt{destination\_task}, the agent's destination if they have just found a task or are headed towards their committed task; \texttt{committed\_task}, the task an agent has committed to; \texttt{angle}, a random walk parameter denoting angle of travel; \texttt{starting\_point}, a random walk parameter tracking where the agent started from; and \texttt{travel\_distance}, the length of the current leg of the random walk.

The agent transition functions $\alpha$ for propagator and follower agents, respectively, use these state variables to implement the desired transitions at each time step. The pseudocode for $\alpha$ for propogator and follower agents can be found in Algorithms \ref{alg:prop-code} and \ref{alg:foll-code}, respectively. The following functions (already described in the steps of PROP) are referenced in the pseudocode: \texttt{propagate()}, the propagator spreads its newest task information $\mathcal{M_T}$ to the other propagators in its influence radius; \texttt{find\_nearby\_task()}, the follower looks at the vertices in its influence radius for a task with nonzero residual demand; \texttt{dir\_from\_propagator()}, the follower looks at the task information $\mathcal{M_T}$ of the propagator at its current vertex, chooses a task according to the probabilities defined earlier, and returns the direction towards that task, or, if $\mathcal{M_T}$ is  empty or zero, returns a random direction.

\begin{algorithm}
\caption{Agent transition function $\alpha$ for a propagator agent with state $s$ at vertex $v$ with coordinates $(x,y)$}\label{euclid}
\begin{algorithmic}[1]
\Procedure{generate\_transition}{local\_vertex\_mapping}
\State $new\_agent\_state \gets s$
\If {$v.\texttt{state.is\_task}$}
\State $new\_agent\_state.\texttt{task\_info}[(x,y)] \gets $
\Indent \State $v.\texttt{state.residual\_demand}$ \EndIndent
\EndIf 
\If {$s.\texttt{propagation\_ctr} \geq s.\texttt{propagation\_rate}$}
\State $self.\texttt{propagate()}$
\State $new\_agent\_state.\texttt{propagation\_ctr} \gets 0$
\Else 
\State $new\_agent\_state.\texttt{propagation\_ctr} \gets$ 
\Indent \State $new\_agent\_state.\texttt{propagation\_ctr} + 1$ \EndIndent
\EndIf 
\State \Return $v.\texttt{state}, new\_agent\_state, \texttt{S}$
\EndProcedure
\end{algorithmic}
\label{alg:prop-code}
\end{algorithm}

\begin{algorithm}
\caption{Agent transition function $\alpha$ for a follower agent with state $s$ at vertex $v$ with coordinates $(x,y)$}\label{euclid}
\begin{algorithmic}[1]
\Procedure{generate\_transition}{local\_vertex\_mapping}
\State $new\_agent\_state \gets s$
\If {$s.\texttt{committed\_task is None}$}
\If {$s.\texttt{destination\_task is None}$}
\If {$self.\texttt{find\_nearby\_task() is not None}$}
\State $new\_agent\_state.\texttt{destination\_task} \gets$ 
\Indent \State $self.\texttt{find\_nearby\_task()}$ \EndIndent
\State \Return $v.\texttt{state}, new\_agent\_state, \texttt{S}$
\Else
\State \Return $v.\texttt{state}, new\_agent\_state, $
\Indent \State $self.\texttt{dir\_from\_propagator()}$ \EndIndent
\EndIf
\Else
\State $\textbf{if } s.\texttt{destination\_task.state.residual\_demand }$
\Indent \State $\texttt{is } 0 \textbf{ then}$ \EndIndent
\Indent \State $new\_agent\_state.\texttt{destination\_task} \gets \texttt{None}$
\State \Return $v.\texttt{state}, new\_agent\_state, \texttt{S}$ \EndIndent
\EndIf
\Indent \If {$s.\texttt{destination\_task} \texttt{ is } v$}
\State $new\_agent\_state.\texttt{committed\_task} \gets $
\Indent \State $s.\texttt{destination\_task}$ \EndIndent
\State $new\_agent\_state.\texttt{destination\_task} \gets \texttt{None}$
\State $new\_vertex\_state \gets v.\texttt{state}$
\State $new\_vertex\_state.\texttt{residual\_demand} \gets$
\Indent \State $ new\_vertex\_state.\texttt{residual\_demand} - 1$ \EndIndent
\State \Return $new\_vertex\_state, new\_agent\_state, \texttt{S}$
\Else
\State \Return $v.\texttt{state}, new\_agent\_state,$
\Indent \State $self.\texttt{dir\_from\_dest(}s.\texttt{destination\_task,}$
\Indent \State {$\texttt{x, y)}$} \EndIndent \EndIndent \EndIf
\EndIndent \Else
\State \Return $v.\texttt{state}, new\_agent\_state, \texttt{S}$
\EndIf
\EndProcedure
\end{algorithmic}
\label{alg:foll-code}
\end{algorithm}

\section{Results} \label{sec:results}
Our algorithms were tested in simulation \cite{sourcecode} using Pygame (see Figure \ref{fig:propsim}) on a grid of size $M=N=50$, with a $3 \times 3$ home area in the center of the grid. Each vertex had an area of $1\text{cm}^2$, meaning that agents moved at $1\text{cm/s}$ (letting time be discretized by $1$s), a speed which simple, low-cost robots are able to move at \cite{rubenstein2012kilobot}. All simulations were run using $100$ task-performing agents---this resulting swarm density was chosen to allow for feasible task discovery time, particularly in the case of RW---and the total task demand summed to $80$. In the trials for the HHTA algorithm, agents had an influence radius of 2. In the trials for the PROP algorithm, propagators had an influence radius of 1 and followers had an influence radius of 2. These smaller influence radii were chosen to keep the algorithms more local.

For each set of trials, we evaluated task completion time, defined as the time necessary for the total residual demand to become $0$. In subsections \ref{subsec:hhta_density} and \ref{subsec:prop_density}, we also measure the average number of messages sent per run per agent. For the HHTA algorithm, whenever a Home agent is notified of a task by a  Recruiting agent, the Recruiting agent's message count is incremented. For the PROP algorithm, the message count is incremented when a propagator shares new task information with one of its neighbors. We do not track the message count for follower agents since it is a negligible portion of total messages.
\subsection{Effects of Task Density on HHTA Performance} \label{subsec:hhta_density}

\begin{figure}
    \centering
    \includegraphics[scale=0.4]{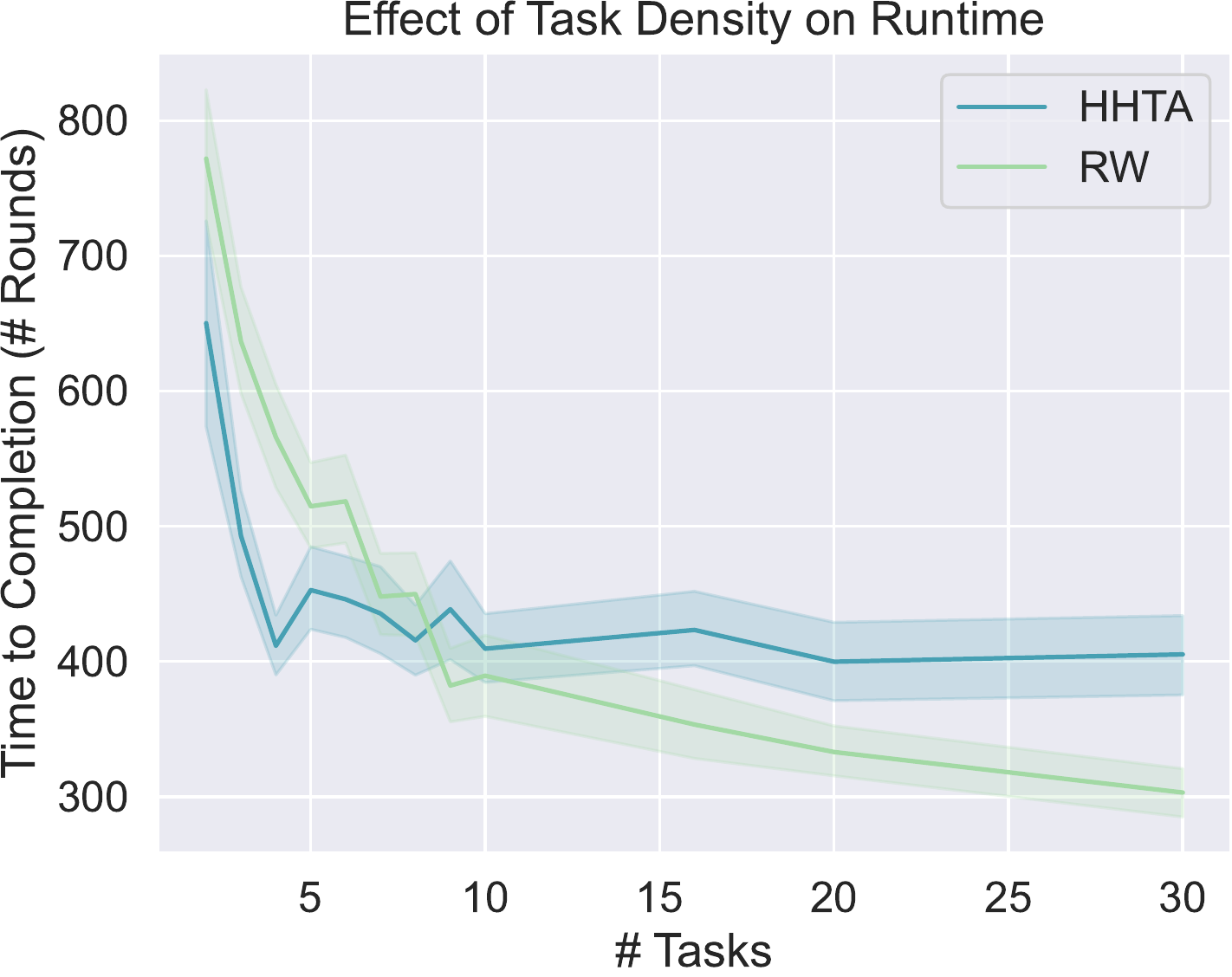}
    \caption{The effect of number of tasks on completion time for HHTA and RW}
    \label{fig:hhta_density_plot}
\end{figure}
\begin{figure}
    \centering
    \includegraphics[scale=0.4]{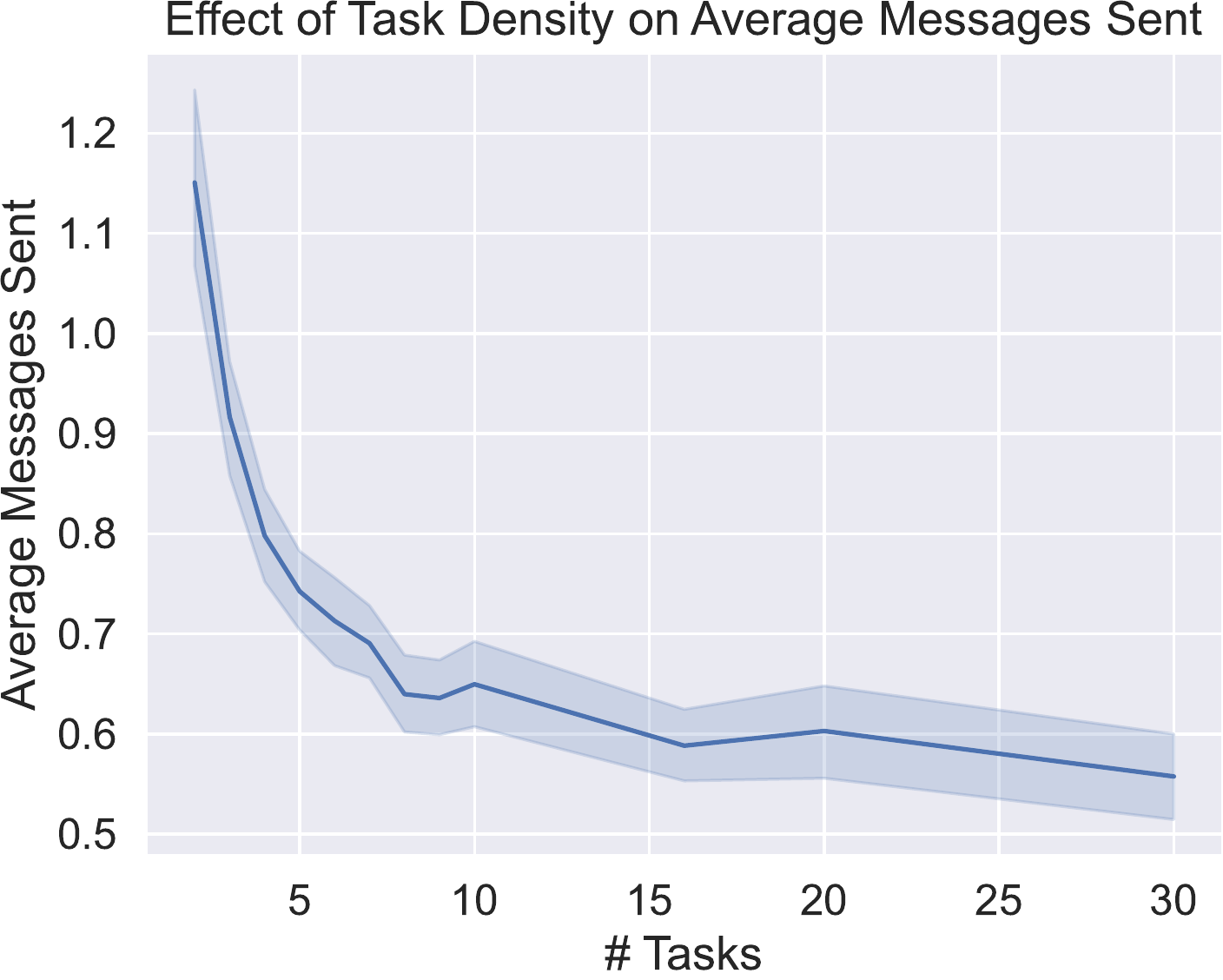}
    \caption{The effect of number of tasks on average messages sent per agent for HHTA}
    \label{fig:hhta_msgs_plot}
\end{figure}

To examine the effects of task density on the HHTA algorithm's performance, we measured task completion time and average number of messages sent per agent for $T \in \{2, 3, 4, 5, 6, 7, 8, 9, 10, 20, 30\}$. For each value of $T$ (the number of tasks), we ran 100 trials with $r_m = \frac{1}{6}$, $P_e = \frac{2}{3}, P_c = \frac{3}{10}$. Figure \ref{fig:hhta_density_plot} shows the resulting average task completion time for varying task densities. The HHTA algorithm outperforms the random walk by about 100 rounds in very sparse task setups when $T \leq 6$ and performs comparably when $7 \leq T \leq 10$, but for denser task setups, the cost of returning to the home nest to recruit others is too high compared to the random walk (Welch's T-test, p=0.05). We can approximate the area covered by detectable tasks as $\frac{T(2I+1)^2}{NM}$, where $(2I+1)^2$ is the size of the influence radius (in reality, the ratio would be a bit smaller as the detectable range for tasks can intersect). This means that for our choice of parameters, the HHTA algorithm outperforms the random walk when about $6\%$ or less of the total task area has an immediately detectable task.


Figure \ref{fig:hhta_msgs_plot} shows the average number of messages sent per agent for the HHTA algorithm. (Note that the random walk algorithm uses no communication). Note that on average, each agent sends less than $1.2$ messages per round using HHTA. Note also that agents send less messages on average as density increases. Since the total task demand is fixed at 80, a larger number of tasks indicates less demand per task on average, making agents in the HHTA algorithm less likely to enter the task state (where messages are sent) and remain in it.

\subsection{Effects of Task Density on PROP Performance} \label{subsec:prop_density}

\begin{figure}
    \centering
    \includegraphics[scale=.4]{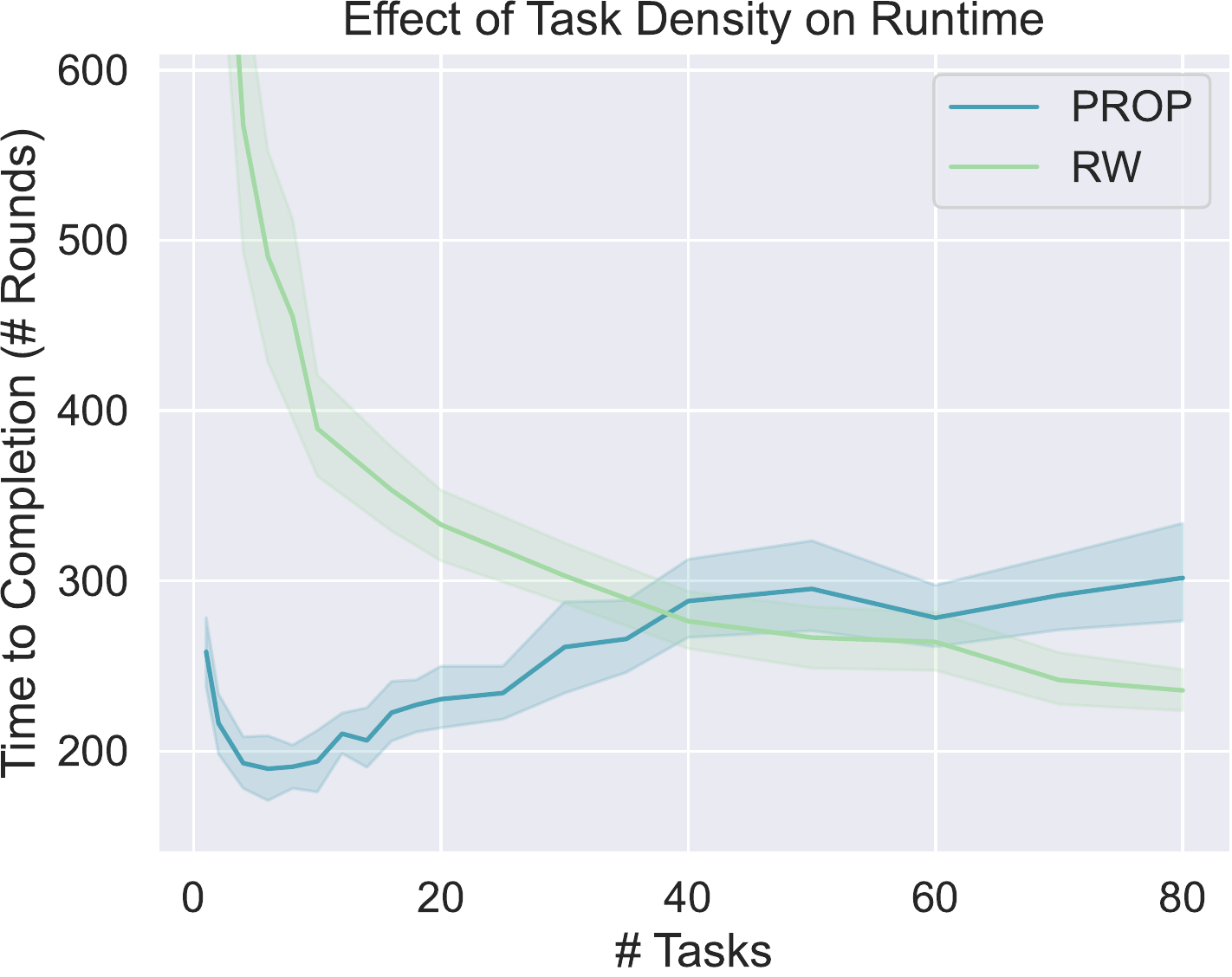}
    \caption{The effect of number of tasks on completion time for PROP and RW}
    \label{fig:prop_sparsenessvruntime}
\end{figure}

To examine the effects of task density on the PROP algorithm's performance, we once again measured task completion time and average number of messages sent per propagator agent for $T \in \{1,2,4,6,8,10,12,14,16,18,20,25,30,35,40,50,60,70,80\}$. For each value of $T$ (the number of tasks), we ran 20 trials with $r_p = 3$ and $d_p = 25$. Figure \ref{fig:prop_sparsenessvruntime} shows the resulting average task completion time for varying task densities. The PROP algorithm outperforms the random walk significantly (Welch's T-test, p=0.05) in sparser task setups ($T \in [1, 25]$), with fewer tasks exaggerating this performance gap nonlinearly. In moderately dense task setups ($T \in [30,60]$), the two algorithms' runtimes are comparable, and in our most dense task setups ($T \in [70, 80]$), the random walk begins to increasingly outperform the PROP algorithm to a significant extent (Welch's T-test, p=0.05). Intuitively, as the density of tasks in the environment increases, follower agents are more likely to find tasks in their influence radius (benefiting RW). Conversely, more tasks means more task information within each propagator agent, overloading and misguiding the follower agents during their decision process (harming PROP). After a certain point, too much propagated information results in that information declining in its specificity and thus usefulness.

\begin{figure}
    \centering
    \includegraphics[scale=.40]{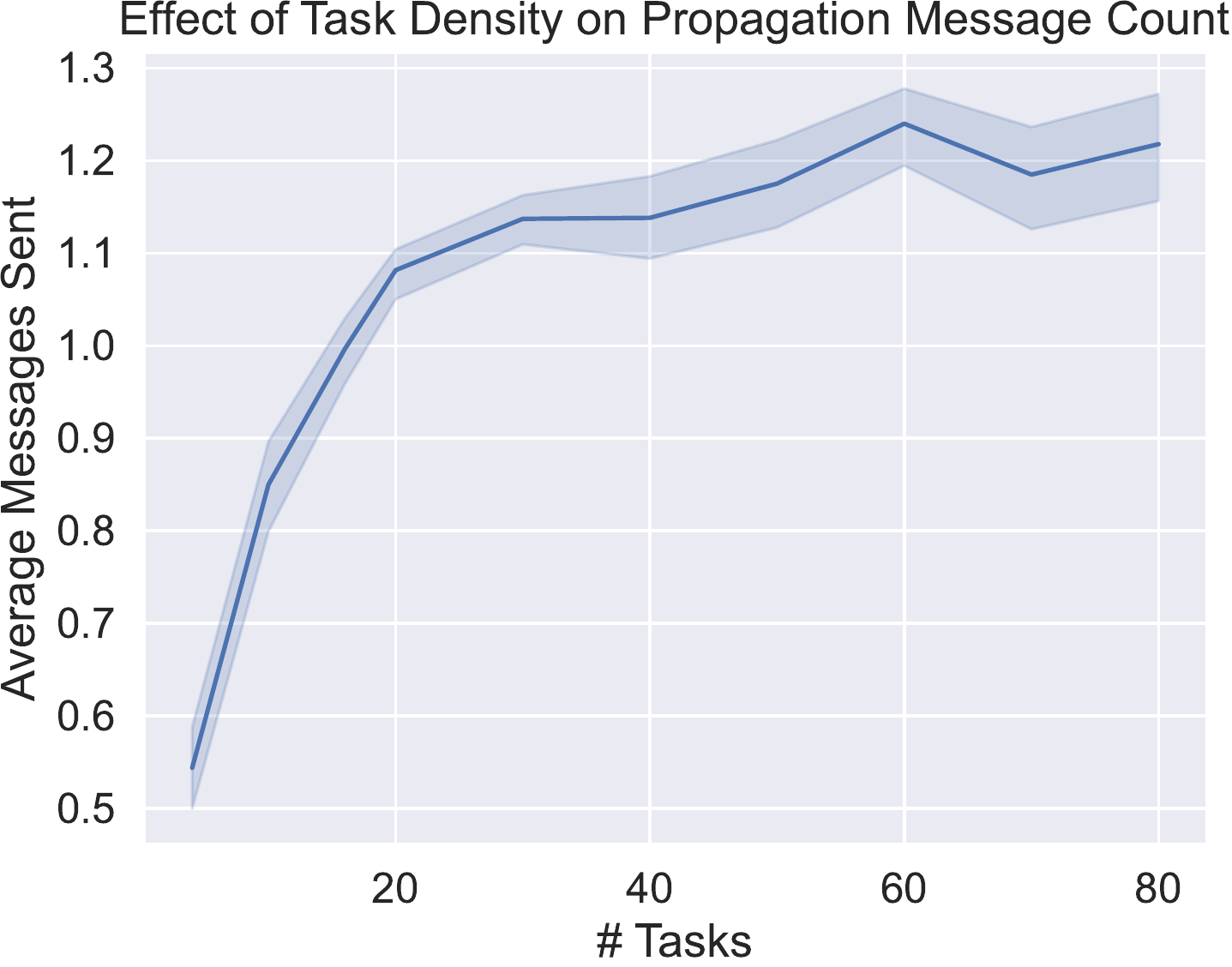}
    \caption{The effect of number of tasks on average messages sent per propagator agent per round (for PROP)}
    \label{fig:prop_sparsenessvmessages}
\end{figure}

Figure \ref{fig:prop_sparsenessvmessages} shows the average number of messages sent per propagator agent per round (for the PROP algorithm). Note that on average, each propagator agent sends less than 1.3 messages to other agents per round; however, given the grid space's size, a large communication cost is still incurred as there are $2,500$ propagator agents. Regarding the effect of task density on these communication costs, agents generally send more messages as the number of tasks increases. When there are more tasks and thus more vertices close to tasks, it takes less time for most of the propagator agents to receive some information initially that they can begin propagating. Additionally, having more tasks means that \textit{some} task's demand gets updated more often, resulting in there being new information (as messages) that needs to be propagated more often. This increasing trend becomes less dramatic at higher task densities, likely due to the fact that with enough tasks, the overlapping propagation radii all cover roughly the same area. As shown in Figure \ref{fig:prop_sparsenessvruntime}, recall that higher task densities result in higher completion times for the PROP algorithm. Therefore, higher task densities do not only result in agents sending more messages per round, but the total number of messages sent over an entire run increases even more dramatically along with the number of tasks.


\subsection{Effects of $P_c$ on HHTA Completion Time for Varying Task Density}
\begin{figure}
    \centering
    \includegraphics[scale=.40]{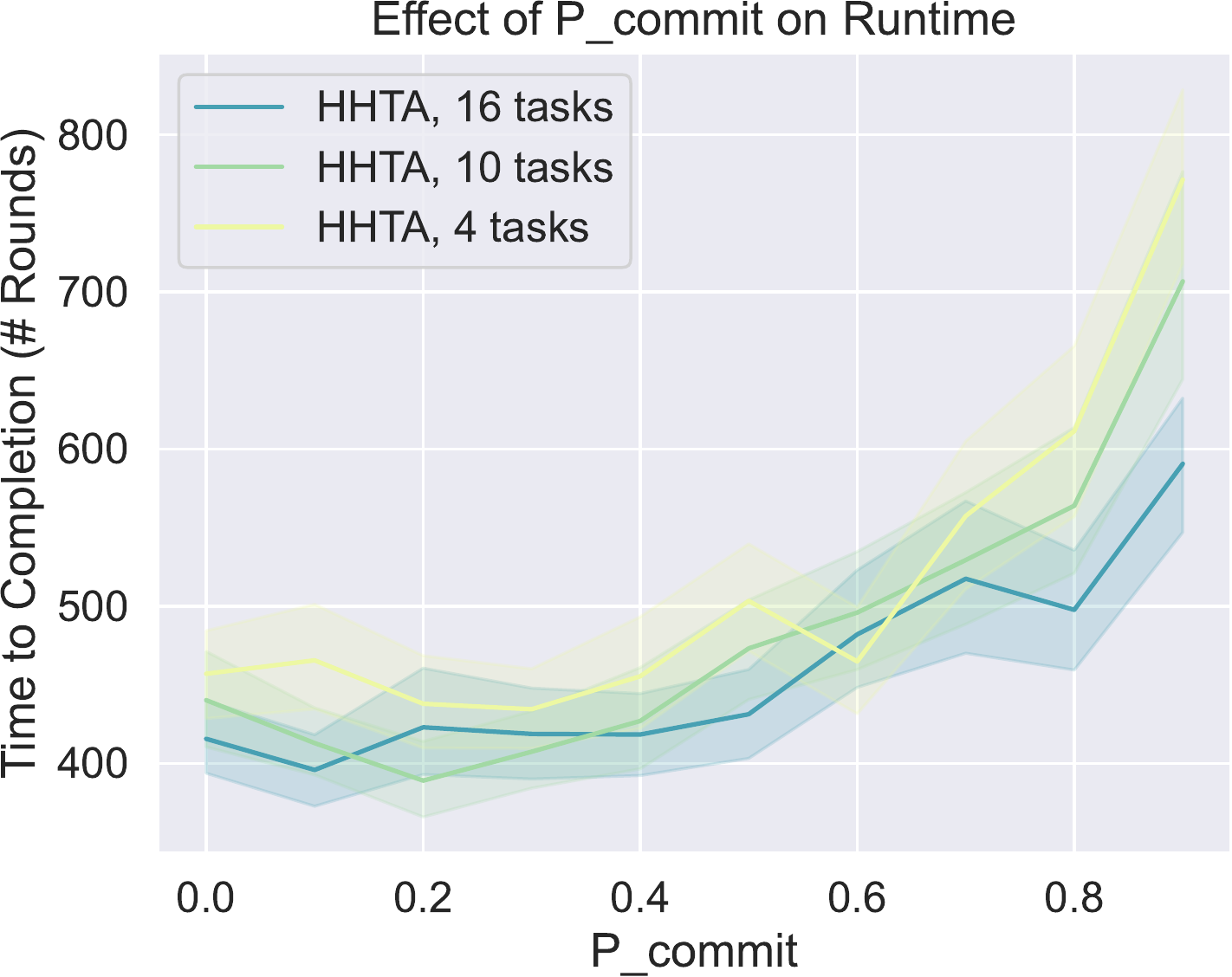}
    \caption{The effect of $P_c$ on HHTA completion time for $\{4, 10, 16\}$ tasks}
    \label{fig:p_commit}
\end{figure}

We explored the effects of varying $P_c$ on completion time for varying $T$. We ran 100 trials for each value of $P_c \in \{\frac{i}{10}, 0 \leq i \leq 9\}$ using $r_m = \frac{1}{6}$ and $P_e = \frac{2}{3}$. The results can be seen in Figure \ref{fig:p_commit}. 

For $P_c \leq 0.7$, there was no significant difference between the HHTA completion time at different task densities (Welch's T-test, p=0.05). However, for $P_c \in \{0.8, 0.9\}$, the completion time for $T=4$ is higher than the completion time for $T=16$ (Welch's T-test, p=0.05). Our results show that only for large $P_c$ do we see a significant difference in performance at different task densities. This makes sense, as a larger proportion of committing agents means agents mostly find tasks by discovering them independently, which is harder in the sparse case. We also note that from $P_c = 0.4$ to $P_c = 0.8$, task completion time follows an increasing trend, indicating that higher recruitment (lower $P_c$) allows agents to complete tasks faster. 

\subsection{Effects of $P_e$ on HHTA Completion Time for Varying Task Density}
\begin{figure}
    \centering
    \includegraphics[scale=.40]{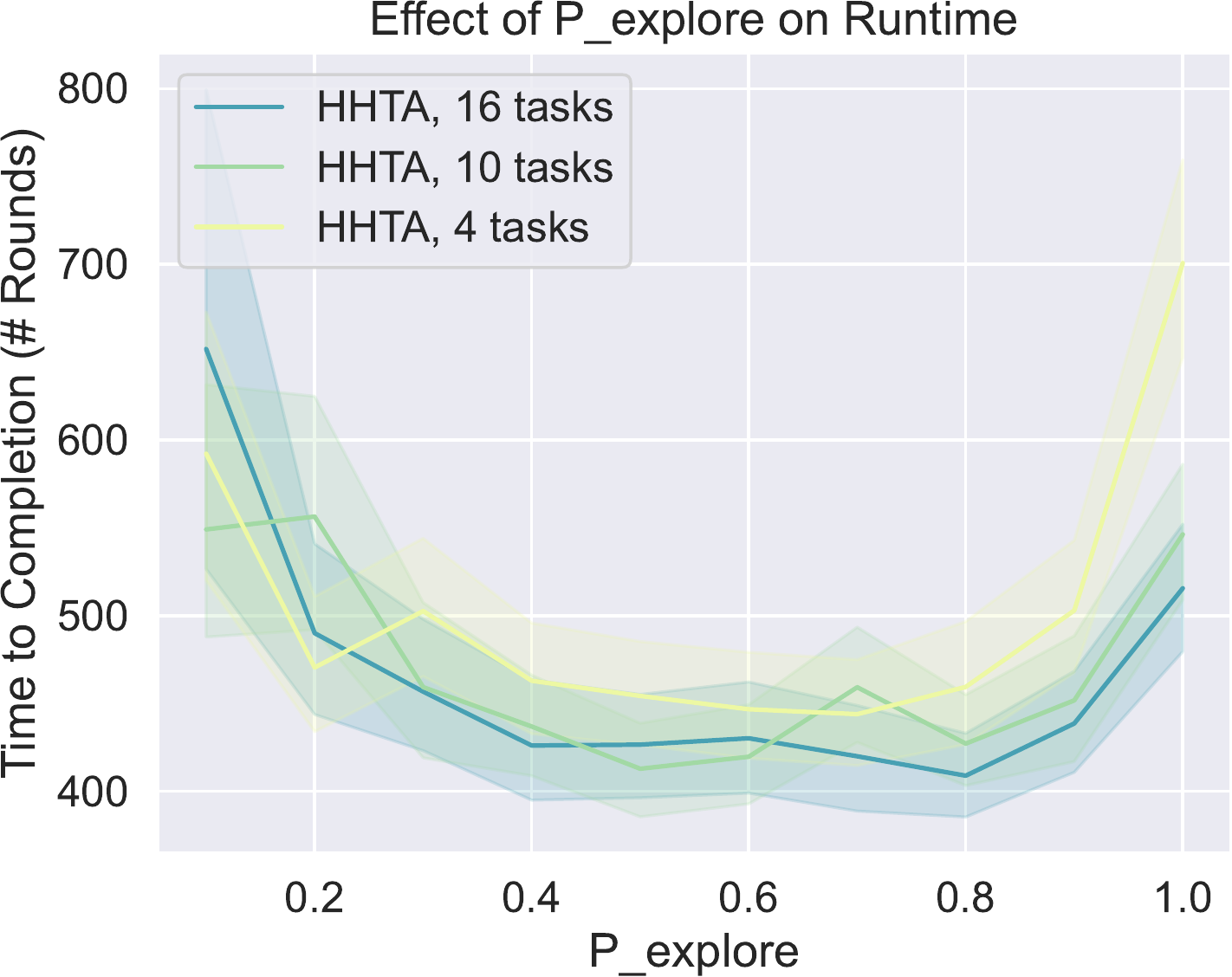}
    \caption{The effect of $P_e$ on HHTA completion time for $\{4, 10, 16\}$ tasks}
    \label{fig:p_e}
\end{figure}

We also explored varying $P_e$, running 100 trials each for $P_e \in \{\frac{i}{10}, 1 \leq i \leq 10\}$, with $T \in {4, 10, 16}$ and using $r_m = \frac{1}{6}$ and $P_c = \frac{3}{10}$. The results can be seen in Figure \ref{fig:p_e}.

For $P_e \leq 0.8$, there was no significant difference between the HHTA completion time at different task densities (Welch's T-test, p=0.05) with the exception of $T = 10$ vs. $T=4$ at $P_e = 0.2$, with $p=0.03$. However, there was a significant difference in completion time between $T=4$ and $T=16$ when $P_e = 1.0$ and $P_e = 0.9$. Our results show that HHTA has a consistent completion time regardless of task density other than for large $P_e \in \{0.9, 1.0\}$, meaning a large majority of agents are exploring (making the algorithm more similar to random walk). When $P_e$ is high, it is harder to complete the sparse problem because exploring is harder in a sparse environment.

Our results also show that a more even balance of $P_e$ (the proportion of Exploring agents) vs. $1-P_e$ (the proportion of Home agents) leads to a faster completion time. When $P_e$ is too low, not enough agents are exploring, making it harder to find tasks. When $P_e$ is too high, not enough agents are available in the home nest to be recruited when tasks are found. 



\subsection{Effects of $d_p$ on PROP Completion Time for Varying Task Density}
\begin{figure}
    \centering
    \includegraphics[scale=.40]{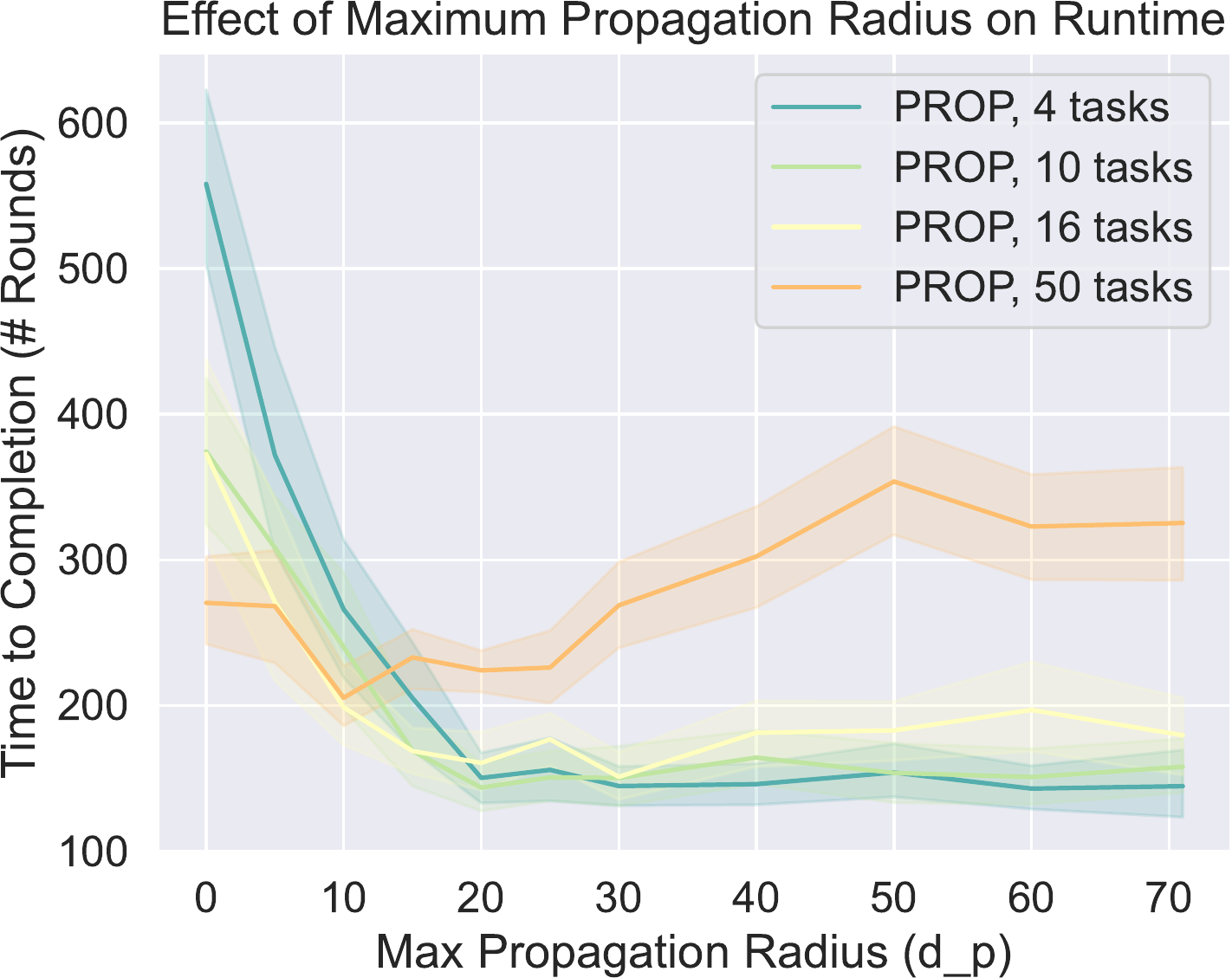}
    \caption{The effect of maximum propagation radius ($d_p$) on PROP completion time for $\{4,10,16,50\}$ tasks}
    \label{fig:prop_propradvruntime}
\end{figure}

We explored the effects of varying $d_p$, the maximum propagation radius, on completion time for varying $T$. We ran 20 trials for each unique pair of $d_p \in \{0,5,10,15,20,25,30,40,50,60,50\sqrt{2}\}$ (note that $d_p=0$ is equivalent to RW, and $d_p=50\sqrt{2}$ means all tasks' information can be propagated over the entire grid space) and $T \in \{4,10,16,50\}$, using $r_p = 3$. The results can be seen in Figure \ref{fig:prop_propradvruntime}.

For reasonably sparse task setups ($T \in \{4,10,16\}$), larger maximum propagation radii correlate with faster runtimes (Welch's T-test, p=0.05) but after a certain point, completion time is mostly unchanged. In contrast, for very dense task setups ($T = 50$), besides a slight improvement in completion time moving from around $d_p = 0$ to $d_p = 10$, larger maximum propagation radii result in slower completion times (Welch's T-test, p=0.05). Increasing $d_p$ results in more propagator agents having more task information, which allows (1) for follower agents to find tasks even if they are far way and (2) for follower agents to leverage this extra task information to prioritize tasks with higher demand. This causes the initial decline in completion time for increasing $d_p$ values. However, if $d_p$ is too large, there is too much information being propagated, diluting the agents' strategy. This adverse effect is likely not seen with sparser environments because even with every single propagator having information about every single task, each mapping of task info is still bounded in size by this smaller number of tasks. It is also reasonable to infer that the turning point in each plot's trend (when completion time either becomes constant or starts increasing) is related to the $d_p$ value at which every propagator agent receives \textit{some} task information.

\subsection{Effects of $r_p$ on PROP Completion Time for Varying Task Density}
\begin{figure}
    \centering
    \includegraphics[scale=.40]{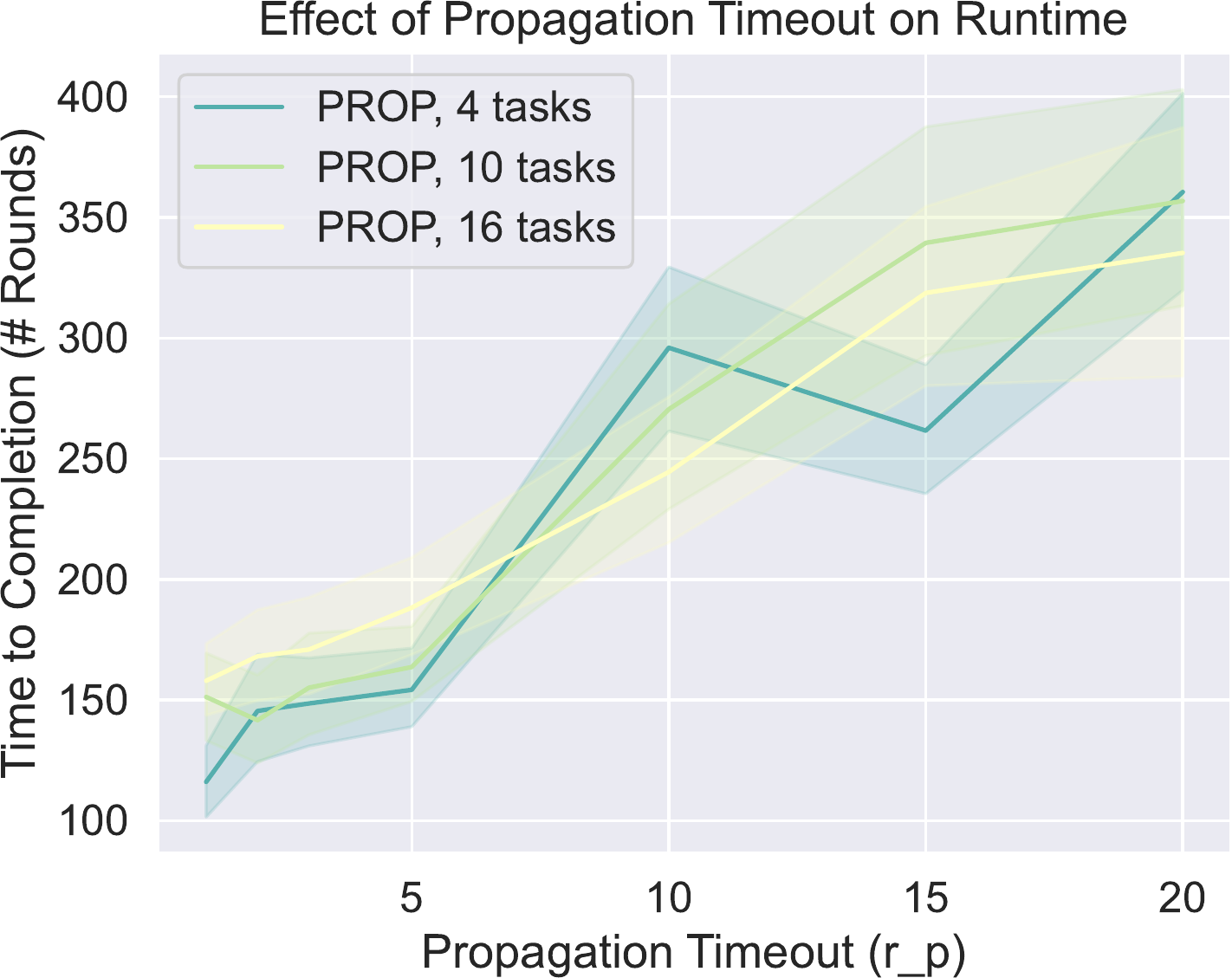}
    \caption{The effect of integer propagation timeout ($r_p$) on PROP completion time for $\{4,10,16\}$ tasks}
    \label{fig:prop_propratevruntime}
\end{figure}

We explored the effects of varying $r_p$, the number of rounds a propagator must wait before sharing new task information with its neighbors, on completion time for varying T. We ran 20 trials for each unique pair of $r_p \in \{1,2,3,5,10,15,20\}$ and $T \in \{4,10,16\}$, using $d_p = 25$. The results can be seen in Figure \ref{fig:prop_propratevruntime}.

There is a clear, mostly linear trend between $r_p$ and completion time, where increasing the propagation timeout results in increasing completion times. The trend is fairly consistent across all distinct task densities that were tested. This relationship between $r_p$ and completion time is to be expected, as smaller $r_p$ means that task information is moved about the environment more quickly, causing the information that is used by follower agents to decide which task to move towards to be more up-to-date. Besides at the very beginning, $r_p$ has no effect on the locations of task information, so none of the adverse phenomena we have seen in which there is ``too much'' propagated information occur when varying $r_p$. It is the same task information, simply better when when the timeout is smaller. It is worth noting, though, that smaller values of $r_p$ involve more message passing.

\section{Discussion} \label{sec:discussion}
Our results demonstrate for both HHTA and PROP that when the total demand for agents is held fixed, task density significantly affects algorithm performance. HHTA performs better than RW when tasks are very sparse, and worse when the number of tasks is high because communicating about individual tasks matters less when there are many of them. RW performs very poorly with sparse tasks because it becomes harder over time for the remaining agents to find tasks. PROP also performs better than RW until the number of tasks is very high, as agents struggle to arrive at tasks when too much task information is being propagated. Though it outperforms RW for sparse tasks, PROP's completion time increases for very sparse tasks ($T \leq 6$). PROP also has a faster completion time and is more distributed than HHTA, but is much more resource and communication intensive, as it requires a propagator agent at every grid cell in order to spread information. 

In relevant task allocation problems such as search-and-rescue or mine detection, the number of tasks in the environment is expected to be sparse, so both algorithms provide a speed-up in completion time compared to the Levy walk. HHTA provides a less agent intensive and less communication intensive approach but requires a central communication location. Contrarily, PROP provides a quicker and more distributed approach for sparse and mid-density environments but is more resource-intensive. Since the Levy flight has been shown to optimize search efficiency and can be observed in many species in nature, it makes sense that for very dense task environments with a low demand per task, the Levy flight outperforms both algorithms. Such environments are a very similar problem to foraging itself. On the other hand, environments with fewer tasks that require more agents benefit more from the coordination and communication of more advanced algorithms.

We also analyzed both algorithms' mechanics individually, showing the importance of recruitment in HHTA as well as the importance of an even balance of Exploring vs. Home agents. For PROP, we showed as expected that generally, higher $d_p$ leads to better performance, though it is more communication-intensive. We also showed that as propagation timeout increases, time to completion increases, since task demands are stale for longer periods of time.

We also note that in extreme parameter settings, HHTA completion time was similar regardless of task density while varying algorithm parameters like $P_c$ and $P_e$. However, this is untrue for PROP, which had a higher completion time for sparser environments at low $d_p$, and a higher completion time for denser environments at high $d_p$. This behavior makes sense because as $d_p$ approaches $0$, PROP reduces to RW, which is similarly affected with a higher completion time for sparse tasks. 

\section{Future Work} \label{sec:future}
Future work could explore experiments in a dynamic setting, where new tasks can appear over time and agents can search for a new task after their existing task is finished. It could also evaluate other environment parameters, such as the ratio of total task demand to total number of agents. A larger such ratio would make the task allocation problem harder to solve, as there are less and less extra agents available to communicate. Another parameter left to be analyzed is swarm density; that is, the ratio of total number of agents to grid size ($M \times N$).

Future work could also combine the strengths of the PROP and HHTA algorithms, where one agent for each task is assigned to propagate by leaving information in the vertex state of nearby vertices or communicating task information directly to any nearby agents like HHTA does. This algorithm would have a much smaller agent cost than the PROP algorithm while still being able to propagate task information. It would also not require a central home nest like the HHTA algorithm does, instead opting to induce agent communication all around the arena. 

Future work could also aim for analytical bounds on the expected task completion time of our two algorithms. Because the algorithms are relatively simple compared to many swarm algorithms, high probability bounds may be possible to obtain.

Lastly, future work could extend our algorithms to the continuous 2D as well as 3D (discrete and continuous) settings, adaptations which our presented theoretical framework is amenable to.



\begin{acks}
Special thanks to Andrea Richa for her insights on the task allocation problem.
\end{acks}



\bibliographystyle{ACM-Reference-Format} 
\bibliography{sample}


\end{document}